\documentclass{aa}

\usepackage{graphicx}
\usepackage{txfonts}
\usepackage{amssymb}
\usepackage{listings}
\usepackage{nameref}
\usepackage{rotating}
\usepackage[utf8]{inputenc}
\usepackage[draft]{hyperref}
\usepackage{booktabs}
\usepackage{color}
\usepackage{adjustbox}
\usepackage[frozencache=true,cachedir=minted-cache]{minted} 

\usepackage{scrextend}
\usepackage{indentfirst}
\usepackage{latexsym}
\usepackage{multirow}
\usepackage{epsfig}
\usepackage{hyperref}
\usepackage[usenames,dvipsnames]{xcolor}
\usepackage{amsfonts}
\usepackage{amsmath}
\usepackage{bm}
\usepackage{algorithm,tabularx}
\usepackage{algpseudocode}
\usepackage{setspace} 
\usepackage{cleveref}

\makeatletter
\newcommand{\multiline}[1]{%
  \begin{tabularx}{\dimexpr\linewidth-\ALG@thistlm}[t]{@{}X@{}}
    #1
  \end{tabularx}
}
\makeatother

\def\cK{ {\cal K} }

\def\cT{ {\cal T} }

\newcommand{\cI}{{\cal I}}

\def\real{\mathbb{R}}

\newcommand{\be}{\begin{equation}}
\newcommand{\ee}{\end{equation}}
\newcommand{\blank}{\bigskip}
\newcommand{\lam}{\lambda}
\newcommand{\Lam}{\Lambda}

\newcommand{\ft}{\footnote}
\newcommand{\code}[1]{\mintinline{Python}{#1}}
\newcommand{\eq}[1]{\begin{align} #1 \end{align}}

\begin{document}

\title{
    Arby - Fast data--driven surrogates
}
\subtitle{}

\author{Aar\'on Villanueva\inst{1}
    \and
    Martin Beroiz\inst{2}
    \and
    Juan Cabral\inst{3, 4}
    \and
    Martín Chalela\inst{4}
    \and
    Mariano Dominguez\inst{4}
}

\institute{
    Facultad de Matem\'atica, Astronom\'ia, F\'isica y Computaci\'on,\\
    Universidad Nacional de C\'ordoba, (5000), C\'ordoba, Argentina
    \and
    LIGO, California Institute of Technology, Pasadena, CA 91125, USA
    \and
    Centro Internacional Franco Argentino de Ciencias de la
    Informaci\'on y de Sistemas (CIFASIS, CONICET--UNR) 
    \and
    Instituto De Astronom\'{\i}a Te\'orica y Experimental -
    Observatorio Astron\'omico C\'ordoba (IATE--OAC--UNC--CONICET)
}

\date{}


\abstract
{The availability of fast to evaluate and reliable predictive models is highly relevant in multi-query scenarios where evaluating some quantities in real, or near-real-time becomes crucial. As a result, reduced-order modelling techniques have gained traction in many areas in recent years.}
{We introduce Arby, an entirely data-driven Python package for building reduced order or surrogate models. In contrast to standard approaches, which involve solving partial differential equations, Arby is entirely data-driven. The package encompasses several tools for building and interacting with surrogate models in a user-friendly manner. Furthermore, fast model evaluations are possible at a minimum computational cost using the surrogate model.}
{The package implements the Reduced Basis approach and the Empirical Interpolation Method along a classic regression stage for surrogate modelling.}
{We illustrate the simplicity in using Arby to build surrogates through a simple toy model: a damped pendulum. Then, for a real case scenario, we use Arby to describe CMB temperature anisotropies power spectra.  On this multi-dimensional setting,  we find that out from an initial set of $80,000$ power spectra solutions with $3,000$ multipole indices each, could be well described at a given tolerance error, using just a subset of $84$ solutions.}
{}

\keywords{
    Reduced Order Modeling --
    Surrogate Models --
    Reduced Basis --
    Empirical Interpolation --
    Python Package.
}

\maketitle


\section{Introduction}
\label{sec:intro}

Several problems arise in observational and theoretical contexts that demand the resolution of computationally intensive differential equations. From structural analysis in engineering to spacetime simulations in astrophysics, rapid and reliable evaluations of solutions to these equations are crucial due to the need for computing in real or near-real time quantities that depend on those solutions. Due to pervasive computational costs led by the inherent complexities present in many problems, achieving fast responses becomes an ubiquitous bottleneck.

As a case study, let us take the example of an ongoing problem in the field of gravitational wave research, namely the template bank problem~(\cite{Field:2012if}). To be able to detect the very faint gravitational wave signals among the noisy background of ground-based interferometers, the LIGO-Virgo collaboration use match filtering against a bank of GW signal templates to maximize the signal-to-noise ratio~(SNR)~(\cite{PhysRevD.49.2658, PhysRevLett.116.241102, Abbott_2020}).

The observed time series is filtered to decide whether a binary coalescence took place, and the parameter estimation allows us to infer the properties of precursors of the remnant, such as masses and spins for binary black hole mergers. 

It is convenient to count with large enough template banks of theoretical waveforms to fill the parameter space of GWs. However, this pose an exceedingly challenging task due to the complexity of solving the Einstein Equations of General Relativity. A single numerically generated GW waveform can take from days to weeks to become available for production~(\cite{Lehner:2014asa}).

A palliative solution is the construction of approximate waveform models, which are more direct to build and deploy than numerical relativity ones. There are several methods to build these approximations. Analytical examples are the Post-Newtonian~(\cite{Blanchet:2006zz}), Effective One Body~(\cite{Damour:2009zoi}) and Phenomenological~(\cite{Sturani_2010, PhysRevLett.113.151101,Khan:2018fmp}) approximations. However, we focus here on a particular set of methods 
that have proven to be very fertile in GW research and produced some of the milestones in waveform modeling in recent years: Reduced Order methods.

Reduced Order Modeling (ROM) is an umbrella term that encompasses a variety of techniques for dimensional reduction, developed to address the problem of complexity in numerical simulations. In particular, Surrogate Models obtained through application of ROM to ground truth solutions are low resolution representations intended to be fast to build and evaluate without compromising accuracy. We take a data-driven approach, i.e. driven {\it only} by data, as opposed to more standard and intrusive ones in which reduction methods are coupled to differential solvers to build solutions~(\cite{quarteroni2015reduced, Jan-RB}).

In waveform modeling, the combination of two ROM methods originally posed for intrusive problems and recreated later for data-driven ones led to significant success in constructing surrogate models for GWs. Those methods are the Reduced Basis (RB)~(\cite{Boyaval20093187, PhysRevLett.106.221102, Field:2012if}) and the Empirical Interpolation (EI)~(\cite{Barrault2004667, sorensen2010}) methods, which we describe in the next section.

The primarily purpose of this work is to disseminate these tools in the astronomy/astrophysics community by introducing a single and user-friendly Python package for data-driven dimensional reduction: Arby.

Arby arises as a response to a lack of well documented, tested and actively maintained code for reduced basis and surrogate modeling in the scientific community while adhering to the data-driven and user-friendliness premise.


\section{Theory Overview}
\label{sec:theory}

This section briefly describes the basics of a reduced-order pipeline for building surrogate models. As we stated above, it merges two main ingredients, the Reduced Basis and the Empirical Interpolation methods, for dimensional reduction of raw data. This pipeline was first introduced in~(\cite{PhysRevX.4.031006}) followed by the construction of several surrogate waveform models based on this method~(\cite{PhysRevLett.115.121102, PhysRevD.96.024058, PhysRevD.95.104023, Varma:2019csw, PhysRevD.99.064045, rifat2019surrogate}). See also~(\cite{tiglio2021reduced}) for a review.

\noindent{\it Representation}.-- We are interested in parametrized scalar (real or complex) functions, solutions or models of the form
\eq{\label{eq:model}
h_\lam(x) := h(\lam; x)\,,
}
where $\lam$ represents the parameter/s of the model and $x$ is the independent variable. Both are real and possibly multidimensional. In physical models, $h_\lam$ can represent parametrized time series with $x$ being the time variable. For convenience, we denote the spaces for $\lam$ and $x$ as the {\it parameter} and {\it physical} domains, respectively.

In the first stage, we look for a low-resolution representation of the model $h_\lam$. The RB method consist in representing a whole set of solutions, usually called the {\it training set}
$$
\cK := \{h_{\lambda_i}\}_{i=1}^N\,,
$$
by linear combinations of basis elements of the form
\be\label{eq:projection}
h_\lambda\approx\sum_{i=1}^n \langle e_i, h_\lambda\rangle e_i\,,
\ee
where
\be\label{eq:inner}
\langle  e_i, h_\lambda \rangle := \int_{\Omega} \bar{e_i}(x) h_\lambda(x) dx
\ee
defines the inner product between training functions, $\bar{e_i}$ is the complex conjugate of $e_i$ --if we deal with complex functions-- and $\Omega$ is the physical domain. The set $\{e_i\}_{i=1}^n$ is called the reduced basis and is composed by a subset of optimally chosen solutions from the training set. The construction of the reduced basis is iterative: at each step of the algorithm, the most dissimilar (orthogonal) element from the training set $\cK$ joins to the current basis, and the process stops when an user-specified tolerance is met. This tolerance is related to the maximum error of the difference between solutions and approximations. In consequence, the different spaces $\mathcal{X}_i(i=1, 2,\ldots)$ spanned by each reduced basis built at each step are nested, i.e. $\mathcal{X}_1\subset \mathcal{X}_2,\ldots$ The addition of a new element to the basis implies a fine tuning of the previous approximation space.

The RB method supposes a compression in the parameter space of solutions. The next step is to achieve compression in time. To this we turn to interpolation replacing the projection-based approach described so far by an interpolation scheme. We pose the problem by looking for an efficient linear interpolation operator $\cI_n$ such that
\be\label{eq:interp}
h_\lambda (x) \approx \cI_n [h_\lambda] (x) = \sum_{i=1}^n C_i(\lam) e_i(x)\,,
\ee
subject to
\be\label{eq:conds}
\cI_n [h_\lam] (X_i) = h_\lam (X_i),\quad i=1,\ldots, n
\ee
for strategically selected nodes $X_i\,(i=1,\ldots,n)$ out from the physical domain. The EI method~(\cite{Maday_2009, Barrault2004667, sorensen2010}) gives us an algorithm for building such interpolant. The Empirical Interpolation (EI) algorithm, as described in~(\cite{PhysRevX.4.031006}), selects iteratively the nodes $\{X_i\}$ from the physical domain following a local optimization criteria~(\cite{Villanueva:2020ixh}).

The EI algorithm receives as {\it unique} input the reduced basis and selects the interpolation nodes for building the interpolant. Note that there is no need of the whole training set $\cK$ since the RB algorithm already did the introspection of it, and we assume that the relevant information about $\cK$ is hard-coded in the reduced basis.

It is possible to show that, under some conditions, the interpolation error is similar to the projection one, which in most applications has exponential decay in the number of basis elements~(see (\cite{tiglio2021reduced}) and citations therein). This leads to an efficient and (in most cases of interest) accurate representation of the training set by means of an empirical interpolation.

To summarizing: first, a reduced basis is built from a training set using the RB method, which leads to the linear representation in~(\ref{eq:projection}). This step completes a compression in the parameter domain. Next, an empirical interpolant is built solely from the reduced basis. This step completes a compression in the physical domain. Finally, we end up with an empirical interpolation~(\ref{eq:interp}) which provides efficient and high-accuracy representations of all functions {\it in} the training set.

\noindent{\it Predictive models}.-- We want our model to represent solutions that are not present in the training set. That is, we look for the predictability. For this, we perform parametric fits at each empirical node along with training values. Let us break it down. 

Let us rewrite the interpolation in~(\ref{eq:interp}) as
\be\label{eq:interpB}
\cI_n [h_\lambda] (x) = \sum_{i=1}^n B_i(x) h_\lam(X_i)\,,
\ee
where
\be\label{eq:defB}
B_i (x) := \sum_{j=1}^n ({\bf V}_n^{-1})_{ji}e_j (x)
\ee
and
$$
 ({\bf V}_n)_{ij}:=e_j(X_i)\,.
$$

Recall the approach is data-driven, so we do not fill the training set with more solutions to approximate newer ones. Instead, we predict them by performing fits along data that we already know, that is, along $h_\lambda(X_i)$ $(i=1,\ldots,n)$. For $1$-D problems ($\lam\in\real$), deciding for problem-agnostic fitting methods that are well-suited for most cases can be a challenging task, not to mention the high dimensional case, which remains an open question~(\cite{tiglio2021reduced}). For a rough classification, we refer to regression and interpolation methods. The former deals with calibrating free parameters of some model by optimizing an error function; the latter, with solving an interpolation problem consisting essentially in solving algebraic systems which are possibly subject to constraints (e.g. Eqs.~(\ref{eq:interp},\ref{eq:conds})). 

Here we address the second approach for parametric fits. The procedure consists in interpolating the values
$$
\{h(\lam_j;X_i)\}_{j=1}^N
$$
along parameter samples for each empirical node $X_i\,,\, i=1,\ldots,n$. As we describe in Section~\ref{sec:details}, as of now, Arby uses {\it splines} for this step, though we expect to generalize this. Once fits are done, we end up with a surrogate model $h_\lam^{surr}(x)$ which can represent and predict $h_\lam$ at {\it any} $\lambda$ in the parameter domain with high accuracy and low computational cost.


\section{Algorithms}
\label{sec:algorithms}

The pipeline for building surrogates described in the previous section is valid in any number of parameter and physical dimensions if we consider arbitrary fittings through parameter space. For surrogate modeling, Arby supports in its present version 1-D parameter and physical domains (real number intervals of the form $[a,b]$ for $a,b\in\real$) and real-valued functions. Again, this restriction is only for building surrogate models. On the other hand, Arby supports multidimensional parameter domains (although still restricted to 1-D domains in the physical dimension) and complex-valued functions for building reduced bases and empirical interpolants.

Below we summarize the algorithm for surrogate modeling. We refer the reader to the Appendices for technical details about the RB and EI algorithms used in intermediate stages.

The inputs are the training set $\cK = \{ h_{\lam_i}\}_{i=1}^N$, the parameter set $\cT := \{\lam_i\}_{i=1}^N$, and the greedy tolerance $\epsilon\in\real$.

\begin{algorithm}[H]
\caption{Surrogate modeling}
\label{alg:surr}
\begin{algorithmic}[1]
\State {\bf Input:} $\cK$, $\cT$, $\epsilon$
\vskip 10pt
\State Build the reduced basis $\{e_i\}_{i=1}^n$ up to tolerance $\epsilon$.
\State Find the empirical nodes $\{X_i\}_{i=1}^n$ and build the interpolant $\cI_n$ by assembling the functions $B_i(x)\,(i=1,\ldots,n)$ (see \ref{eq:defB}).
\For{$i=1 \to n$}
\State \multiline{
  Build a continuous function $h_i^{fit}(\lam)$ by

  doing fits along values $\{h(\lam;X_i)\}_{\lam\in\cT}$.
}
\EndFor
\State Assemble the surrogate:

$h^{surr} (\lam; x) := \sum_{i=1}^n B_i(x) h_i^{fit}(\lam)$
\vskip 10pt
\State {\bf Output:} surrogate model $h^{surr} (\lam; x)$
\end{algorithmic}
\end{algorithm}

Let's make some remarks on Alg.~\ref{alg:surr}.

\begin{itemize}
\item The training set $\cK$ is built from a discretization $\cT$ of the parameter domain. In the current version of Arby, $\cT$ is a discretized real interval $\cT := \{\lam_1,\ldots,\lam_N\}$.

\item The RB algorithm used for building the reduced basis is fully described in Alg.~\ref{alg:Greedy}, see the Appendices. It selects from $\cT$ $n$ points $\lambda_i=\Lambda_i\,(i=1,\ldots,n)$, called the {\it greedy} points, labeling those functions in the training set that conform the reduced basis. For conditioning purposes, the basis is orthonormalized at each step, so the algorithm's final output is a set of orthonormal basis elements along with the set of greedy points. So we use the term {\it reduced basis} interchangeably for both, the basis conformed by greedy solutions and its orthonormal version, due to its equivalence (they span the same space).

The number $n$ depends on the greedy tolerance $\epsilon$. In Arby we must specify a discretization of the physical domain $\Omega$ so as to be able to do integrals (see (\ref{eq:inner})). In this context, $\Omega$ is a real interval $[x_a, x_b]$ and Arby must receive as input an equispaced discretization of it.

\item Step 3 implements the EI algorithm described in Alg.~\ref{alg:EIM}, see the Appendices. It receives the reduced basis as unique input and finds $n$ empirical nodes $X_i\,(i=1,\ldots,n)$ to build the interpolant. In practice, the interpolant is specified by assembling the $n$ functions $B_i(x)$ defined in Eqs.~(\ref{eq:interpB},\ref{eq:defB}).

\item To achieve predictability Steps 4-6 perform parametric fits along training values for each empirical node $X_i$. Let's illustrate this by looking at the first iteration of the loop in Steps 4-6. For the first node $X_1$ collect all values
$$
\{h(\lam_1; X_1),h(\lam_2; X_1),\ldots,h(\lam_N; X_1)\}
$$
 and perform a fit along them. This results in a function $h^{fit}_1(\lam)$ that is continuous in the interval $[\lam_1, \lam_N]$. This is repeated $n$ times for each empirical node. The resulted functions $h(\lam)_i^{fit}$ constitute along the reduced basis the building blocks for the final surrogate assembly. The current Arby version implements splines~(\cite{ahlberg1967}) for parametric fits, i.e., piecewise polynomial interpolation of some degree arbitrarily set by the user.

\item From \cref{eq:interpB,eq:defB} the empirical interpolant is defined through the $n$ functions $B_i(x)$. They comprise all the RB-EI information. Combining the functions $B_i(x)$ with the fits $h^{fit}_i(\lam)$ built on previous steps, Step 7 leads to the desired surrogate $h^{surr}(\lam;x)$ which is continuous in $\lam$ inside the real interval $[\lam_1, \lam_N]$.

\end{itemize}

\subsection{Related works}
\label{sec:cmp}

A previous implementation of the RB-EI approach is GreedyCpp~\footnote{\url{ttps://bitbucket.org/sfield83/greedycpp/}}, an MPI/OpenMP parallel code written in C++~(\cite{Antil:2018efn}). Although it is not designed for building surrogates and training sets have to be loaded at runtime, it allows for building reduced bases, empirical interpolants and reduced-order quadratures. Another example is ROMPy~\footnote{\url{https://bitbucket.org/chadgalley/rompy/}}, a previous attempt written in pure Python which supports surrogate modeling.

Other ROM implementations in the Python ecosystem are not fully data-driven. Typically they are weakly or strongly coupled to solvers for differential equations. Mature examples are PyMOR~(\cite{milk2016pymor}) and RBniCS~(\cite{Ballarin2015RBniCSR}). The latter built on top of the FEniCS~\cite{fenics} library for differential equations, and the former allows for coupling with external solvers.


\section{Arby}\label{sec:details}

Arby is a Python package for data-driven surrogate modeling satisfying standard software compliance along quality assurance (see Section~\ref{sec:quality}). It allows the user for 
building reduced basis and empirical interpolants at any number of parameter dimensions. At the current release, Arby builds surrogate models for 1-D domains.

\subsection{Implementation} 
\label{sec:implementation}

Integrals and inner products (see \ref{eq:inner}) must be discretized in implementing Alg.~\ref{alg:surr}. For this, the physical interval $\Omega$ is sampled in $L$ equispaced points $\{x_1,\ldots,x_L\}$ to define a discrete inner product between two functions,
$$
\langle h_1, h_2 \rangle_d := \sum_{i=1}^L \bar{h_1}(x_i) h_2(x_i) \omega_i\,.
$$
The bar represents complex conjugation in case of $h$ being complex. The $\omega_i$'s are $L$ positive real values called {\it weights}. Weights $\{\omega_i\}$ and sample points $\{x_i\}$ constitute a {\it quadrature rule}. Arby uses quadrature rules to compute integrals.

 \subsection{Public API} 
 \label{sec:api}
 
\noindent{\it Classes}.-- The main class in Arby is \code{ReducedOrderModel}. It implements Alg.~\ref{alg:surr} for surrogate modeling. There are three basic inputs for this class:

\begin{itemize}
\item \code{training_set}: 2-D array storing training functions (it corresponds to $\cK$ in Alg.~\ref{alg:surr});
\item \code{physical_points}: 1-D array storing an equispaced discretization of the physical interval $\Omega$ (see \ref{eq:inner});
\item \code{parameter_points}: 1-D array storing the parameter points ($\cT$ in Alg.~\ref{alg:surr});
\end{itemize}

These inputs represent the minimum and indispensable for building surrogates. Optional parameters are:

\begin{itemize}
\item \code{integration_rule}: a string specifying the quadrature rule;
\item \code{greedy_tol}: the greedy tolerance $\epsilon$ for the reduced basis, and
\item \code{poly_deg}: the polynomial degree of splines.
\end{itemize}

These parameters can be tuned for controlling the model accuracy. See the Arby documentation~\footnote{\url{https://arby.readthedocs.io}} for a thorough tutorial on this.

Once a \code{ReducedOrderModel} object is created, the construction and evaluation phases of the surrogate are unified in a single call: just invoke the surrogate at some parameter/s to obtain a prediction.

In principle, the first call of the surrogate is more expensive than subsequent ones, but this is not a problem given that it is done only once. The algorithm design allows upcoming calls to be of fast deployment. By doing so, the process was split in {\it offline-online} stages~(\cite{PhysRevX.4.031006}). Thus, the offline stage corresponds to a (possibly) expensive first building; the online one corresponds to fast model evaluations.

There is a class to compute inner products and integrals: the \code{Integration} class. It receives a 1-D array corresponding to the physical points and the integration rule. Available rules are Riemann (default), Trapezoidal and Euclidean. The former two works for continuous data; the latter for discrete data reduces to discrete inner products. Once a rule is specified, the following methods are unlocked: \code{integral}, \code{dot}, \code{norm} and \code{normalize}. They compute integrals, inner products, norms and normalizations, respectively.

The \code{Basis} class encompasses data utilities for handling arbitrary bases, whether they are reduced bases or user-specified ones. The \code{Basis} class receives as input a basis and an \code{Integration} object. Available methods are: \code{interpolate}, \code{project} and \code{projection_error}. The former two take input arrays and interpolate/project them in a finite-dimensional space spanned by the basis. The interpolation is empirical, i.e. following the EI directives. The projection is simply an orthogonal projection onto the basis. The \code{projection_error} method computes squared projection errors due to projecting arrays onto the basis.

Auxiliary classes are \code{RB} and \code{EIM}. They are containers for RB/EIM information.

\noindent{\it Functions}.-- The main function in Arby is \code{reduced_basis}. It builds a reduced basis giving as input a training set and physical points. This function takes full advantage of the hierarchical feature of the RB algorithm, leading to constant complexity with the number of basis elements~(\cite{tiglio2021reduced}). For conditioning purposes, there are two patterns related to the normalization of the training set which lead to two different implementations of the greedy algorithm.

There is a function for Gram-Schmidt orthonormalization called \code{gram_schmidt}. It implements an Iterative Gram-Schmidt algorithm~(\cite{HoffmannIMGS}) to orthonormalize a set of linearly independent arrays. Internally, this algorithm builds the reduced bases.
\blank

\subsection{Benchmarks} 
\label{sec:benchmarks}

It is interesting within the work to present a performance analysis of the the most important routine of the project: \code{reduce_basis}.

From a theoretical point of view, we expect an overall complexity of $O(nNL)$, where $N, L$ are the number of training and physical samples, respectively, and $n$ is the number of basis elements. Since the training set size is fixed, the addition of one element to the basis at some stage of the greedy loop carries a constant computational cost. If data is amenable for dimensional reduction, we hope $n\ll N$. On the other side, in the worst case scenario we expect $n\simeq N$. This happens if there is almost no redundancy in data to exploit, for example random data. In this case the cost grows as $O(LN^2)$.

Besides theoretical assumptions, we intend to empirically know how the different parameters of \code{reduce_basis} function influence the general performance of the algorithm and if this adjusts to the theoretical estimates.
These parameters are

\begin{itemize}
    \item \code{integration_rule} -- The quadrature rule to define an integration scheme, can be: \textit{riemann}, \textit{trapezoidal} and \textit{euclidean}.

    \item \code{normalize} -- \textit{True} if training data must be normalized before training or \textit{False} otherwise.

    \item \code{greedy_tol} -- The greedy tolerance as a stopping condition for the reduced basis. This allows to control the representation accuracy of the basis. We choose to test two different tolerances, $10^{-14}$ and $10^{-12}$.

    \item \code{training_set} -- The training data as a 2-D array. We tested on square random arrays with sizes $11 \times 11$ and $101 \times 101$.

    \item \code{physical_points} -- Physical points for quadrature rules. Must match the number of columns of training set.

\end{itemize}

With these parameters, we simulated 100 training sets for each one of the 24 possible combinations, giving a total of 24000 test cases. The benchmark was then executed on a computer with the following specifications:

\begin{itemize}
    \item CPU -- 4 x 2.4 GHz AMD Opteron(tm) Processor 6282 SE.
    \item RAM -- 251GB DDR3L.
    \item OS -- CentOS 7 Linux 3.10.0-514.el7.x86\_64
    \item Software -- Python 3.9.0.final.0 (64 bit), NumPy 1.21.1 and SciPy 1.7.0
\end{itemize}

\begin{figure}[h!]
\begin{center}
\includegraphics[width=.8\linewidth]{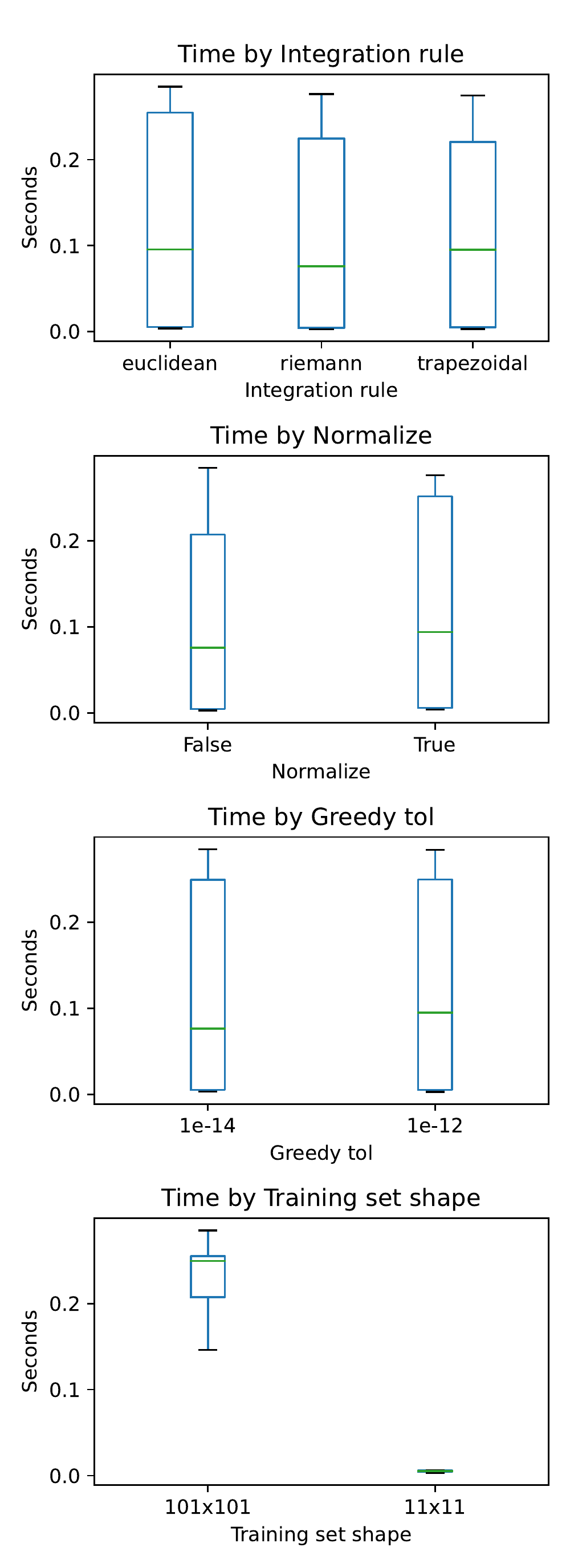}
\caption{
Results of the benchmark on 24000 test cases varying the values of
\code{integration_rule}, \code{normalize}, \code{greedy_tol} and \code{training_set_shape}. In all cases, the horizontal axis represents the different
parameter values, and the vertical axis represents the
execution time in seconds. We can see that the algorithm increases execution time as the
size of the \code{training_set_shape} increases, while in all other cases the times remain relatively unchanged.}
\label{fig:aparams_bench}
\end{center}
\end{figure}

The results are presented in Figure~\ref{fig:aparams_bench}. As we can anticipate, the size of the training set is the most important factor impacting the execution times. All other parameters, out of \code{greedy_tol}, are mainly used for initial configurations outside the greedy loop, hence used only once.

To further explore the relationship between \code{training_set} sizes and execution time, we run a second benchmark leaving fixed \code{normalize=False}, \code{integration_rule='riemman'} and \code{greedy_tol=1e-12}. We generate 100 random \code{training_set} by varying the size between $ 11 \times 11 $ and $ 300 \times 300 $, giving a total of $28,900$ test samples.

\begin{figure}[h!]
\begin{center}
\includegraphics[width=1.\linewidth]{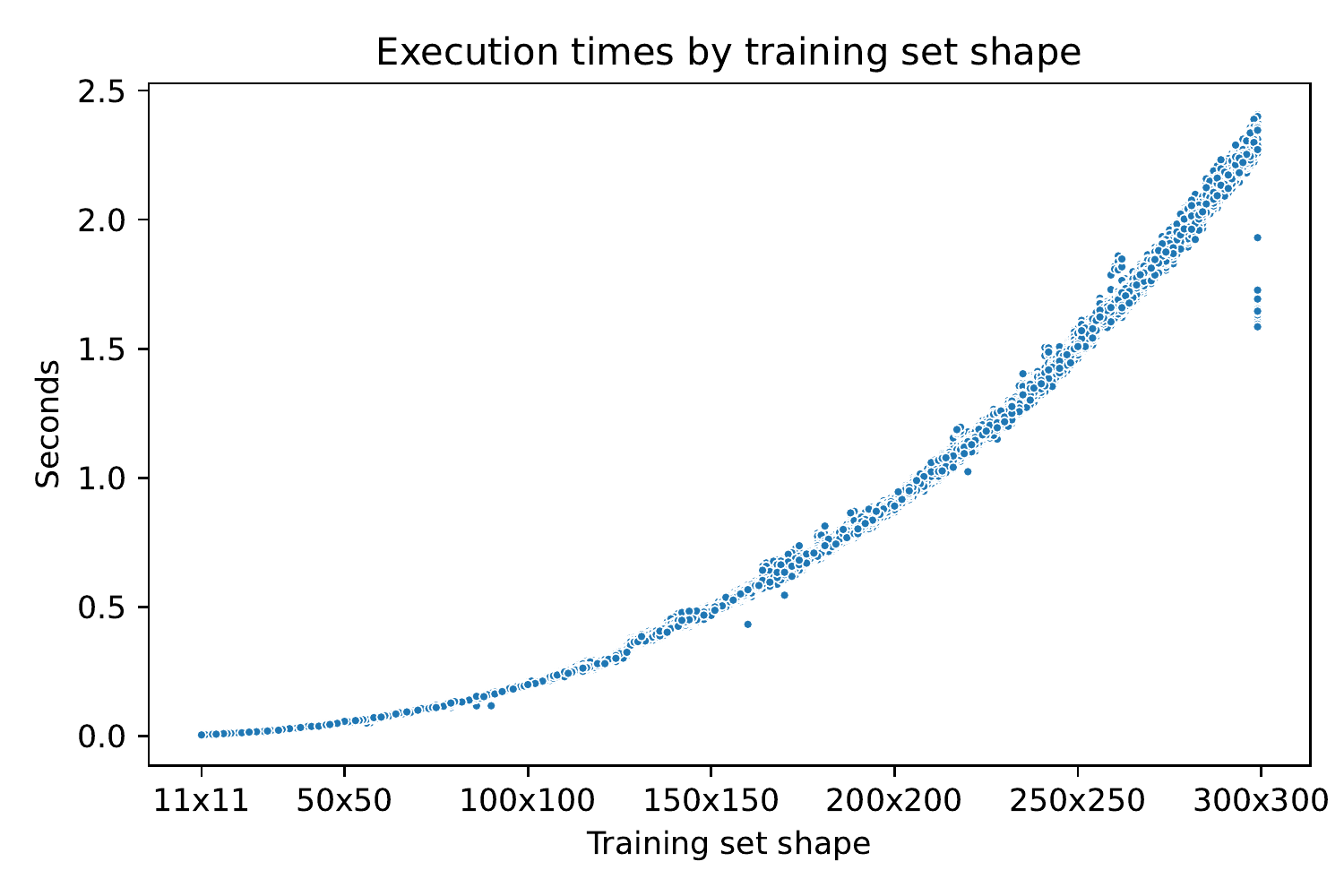}
\caption{
Measured times for $28,900$ test samples of \code{training_set} between $ 11 \times 11 $ and $ 300 \times 300 $. The samples keep the \code{integration_rule}, \code{greedy_tool} and \code{normalize} fixed at \textit{riemman}, $10^{-12}$ and \textit{False}, respectively.}
\label{fig:dbenchmark}
\end{center}
\end{figure}

In this case, the results (Figure~\ref{fig:dbenchmark}) show the anticipated behavior: in front of random others the cost has a growth $O(LN^2)$ as we increase the size of \code{training_set}.

For more details the entire benchmark dataset can be found in \footnote{\url{https://zenodo.org/record/5139187\#.YP-IAXVKhhE} \citep{villanueva_aaron_2021_5139187}}.

\subsection{Quality assurance} 
\label{sec:quality}

To ensure the proper software quality of Arby, we provide standard quantitative and qualitative metrics, in particular i) {\it unit testing} and ii) {\it code-coverage}, and adhere to the PEP $8$ style guide\citep{pep8} throughout the entire project.

\begin{enumerate}
\item Unit testing:

Its purpose is to ensure that the individual software components work as expected~\citep{10.1109/FOSE.2007.26}.
\item Code-coverage:

It measures the amount of code covered by the unit test suite, expressed as a percentage of executed sentences\citep{Miller1963SystematicMA}. By providing comprehensive code-coverage we ensure code validation, expand the ability and efficiency of error handling, and increase confidence in the code.
\end{enumerate}

Arby currently uses pytest \citep{okken2017python} and Coverage.py \footnote{\url{https://coverage.readthedocs.io}} for unit testing and coverage, respectively, completing up to $99\%$ of code-coverage using Python versions $3.6$, $3.7$, $3.8$ and $3.9$.

The {\it PEP $8$ - Style Guide for Python Code} \citep{pep8} is one of a series of guidelines and practices on how to write Python code to improve code readability and consistency. There are several of PEP's (Python Enhancement Proposals), including PEP 8. The latter has recommendations for code layout, whitespaces, comments, naming conventions and programming recommendations.  In addition, there are tools, called {\it linters}, that can be used, in particular, to automate compliance with PEP 8; Arby currently uses {\it flake8}~\footnote{\url{https://pypi.org/project/flake8/}}, which checks for any deviation in code style.

Finally, the entire source code is MIT-licensed \citep{MITlicense} and is publicly available from its GitHub repository\footnote{\url{https://github.com/aaronuv/arby}}. All versions committed to this code are automatically tested in a continuous-integration service using Travis CI\footnote{\url{https://travis-ci.com/github/aaronuv/arby}} and GitHub Actions \footnote{\url{https://github.com/features/actions}}. Documentation is automatically built from the repository and made public in the {\it read-the-docs}~service at ~\footnote{\url{https://arby.readthedocs.io/en/latest/}}.

Arby is built on top of the Python scientific stack: {\it Numpy} \citep{Walt2011TheNA} to perform efficient numerical linear algebra operations; {\it Scipy} \citep{scipy}, used in the current release for splines interpolation.

The Arby package is available for installation on the {\it Python-Package-Index} (PyPI) and can be installed using the command \code{pip install arby}.


\section{Toy model: a damped pendulum}\label{sec:examples}

We illustrate the construction of surrogate models applying Arby to a classical problem in physics: the damped pendulum. This system is a simple pendulum of given longitude subject to gravity and a dissipative force such as friction, allowing for the pendulum oscillations to damp at long times. We encode the generic dynamics of this system in the ordinary differential equation (ODE)
\be\label{eq:pend}
\ddot\theta = -b \dot\theta - \lam \sin(\theta)\,,
\ee

where $\theta$ represents the time-dependent angle of the pendulum with respect to the equilibrium axis and dots represent time differentiation. The symbols $b$ and $\lam$ represent friction and gravity strength per unit length, respectively, at fixed values of the pendulum's longitude. Time units do not play any role in this example, so for practical purposes, we choose it adimensional. This election makes the two parameters of the model, $b$ and $\lam$, also adimensional. In order to make the model one-dimensional and being able to apply Arby for surrogate modeling, fix $b$ to a convenient value so as to cover the variation of solutions in the selected time range widely. Fixed $b$, our parametrized model consist on damped oscillations $\theta_\lam(t)$, being time $t$ the physical variable and $\lam$ the parameter.

\begin{figure}[H]
\begin{center}
\includegraphics[width=1.\linewidth]{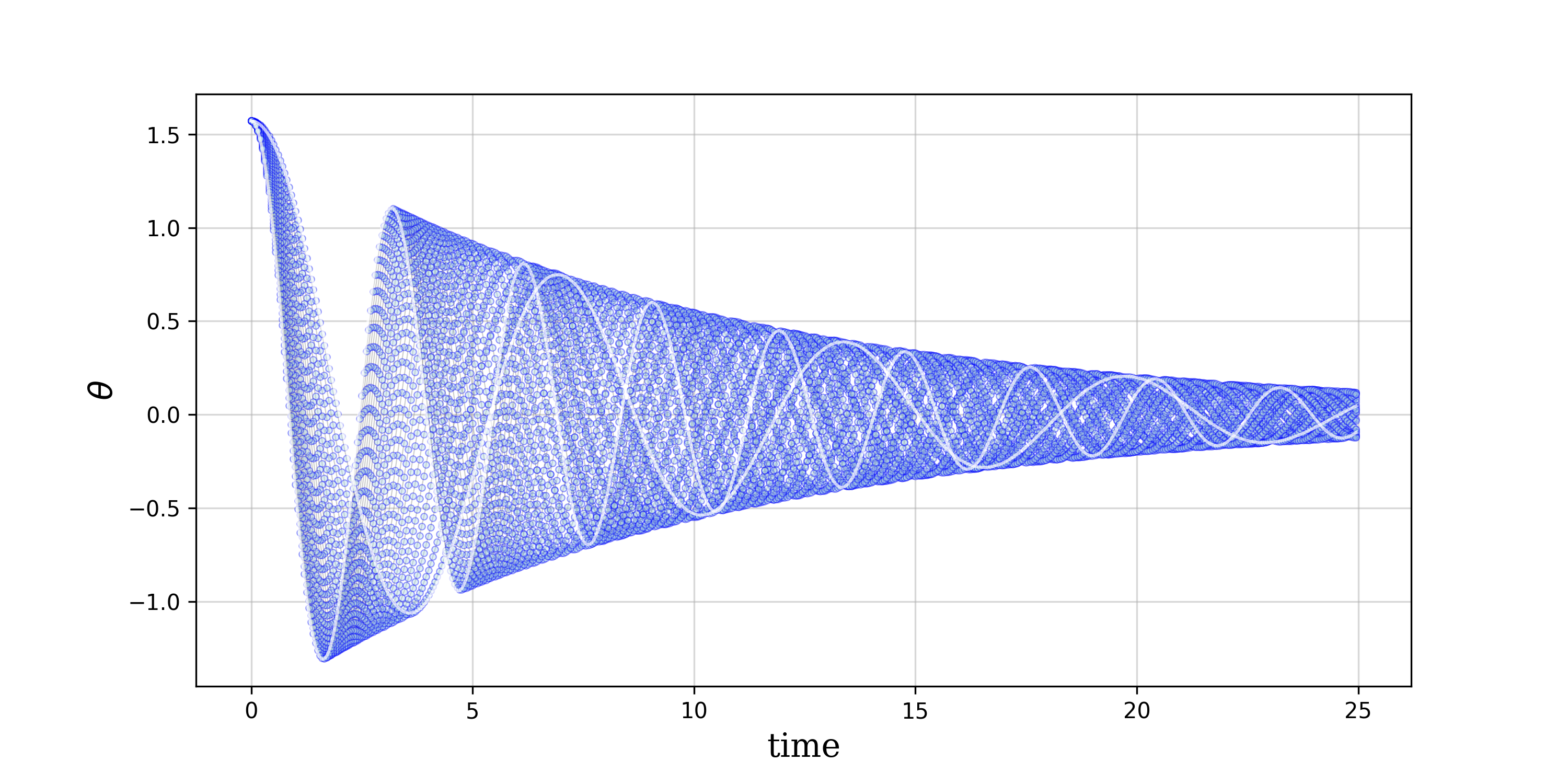}
\caption{Graphical illustration of the training set. We plot a subset of training solutions.}
\label{fig:pend_sols}
\end{center}
\end{figure}

We must solve numerically the ODE in (\ref{eq:pend}) in order to generate the training set of solutions to feed Arby. To this, we set $b=0.2$ and choose intervals for physical and parameter ranges as $t\in[0, 50]$ and $\lam\in[1,5]$. The initial conditions are set to $(\theta, \omega) = (\pi/2, 0)$, where $\omega := \dot\theta$, meaning the pendulum departs from rest at $\theta = \pi/2$ and falls under the action of gravity. We generate a training set of solutions using an ODE solver from the Scipy Python package~\citep{scipy}. We discretize the parameter and time domains in $101$ and $1,001$ equispaced points respectively, and generate $101$ solutions using the same initial conditions\ft{The code for this example can be found in~\url{https://arby.readthedocs.io/en/latest/}}. See Fig.~\ref{fig:pend_sols}.

To build a surrogate for solutions to the pendulum equation, we invoke the \code{ReducedOrderModel} class and feed it with the three main parameters: training set, physical points and parameter points which we denote \code{training}, \code{time} and \code{param}, respectively. 

In addition, we modify default values of optional class parameters to increase the surrogate's quality:
 
\begin{minted}[breaklines, frame=lines, framesep=2mm]{pycon}
>>> from arby import ReducedOrderModel as ROM

# set the greedy tolerance to 1e-14 and
# splines degree to 5 and create a model
>>> pendulum = ROM(training, time, param, greedy_tol=1e-14, poly_deg=5)
\end{minted}

Once the model for the pendulum is created the next step is to build/evaluate the surrogate. For that, we just call it.

\begin{minted}[breaklines, frame=lines, framesep=2mm,]{pycon}
# define the parameter `par` for surrogate evaluation
>>> par = 2.
# evalute
>>> pendulum.surrogate(par)
array([1.57079633, 1.5683046 , 1.56086267, ..., 0.00993474, 0.00975876, 0.009536])
\end{minted}

In order to test the surrogate's accuracy, we build a test set composed of $1,001$ solutions ($10\times$ denser than training set) for the same parameter and physical domains. It means the test set contains the training set plus several solutions not used in the training stage.

We use the $L_2$ norm to compute the relative error between surrogate and ground truth,

\be\label{eq:l2}
e(\lam) := \frac{\| \theta_\lambda^{surr} - \theta_\lambda\|}{\|\theta_\lam\|}
\ee
where
$$
\|\theta_\lam\| := \left( \int_{[0,50]}|\theta_\lambda(t)|^2 dt\right) ^{\frac{1}{2}}\,.
$$

This metric allows us to quantify how well the surrogate globally matches the ground truth model at some parameter value. Furthermore, we compute these errors not only for the surrogate model based on $101$ training parameters but for a set of surrogates built upon different discretizations, starting from very sparse ones until reaching the discretization of $101$ training parameters. With the integration tools available in Arby, these computations become straightforward.

\begin{figure}[H]
\begin{center}
\includegraphics[width=1.\linewidth]{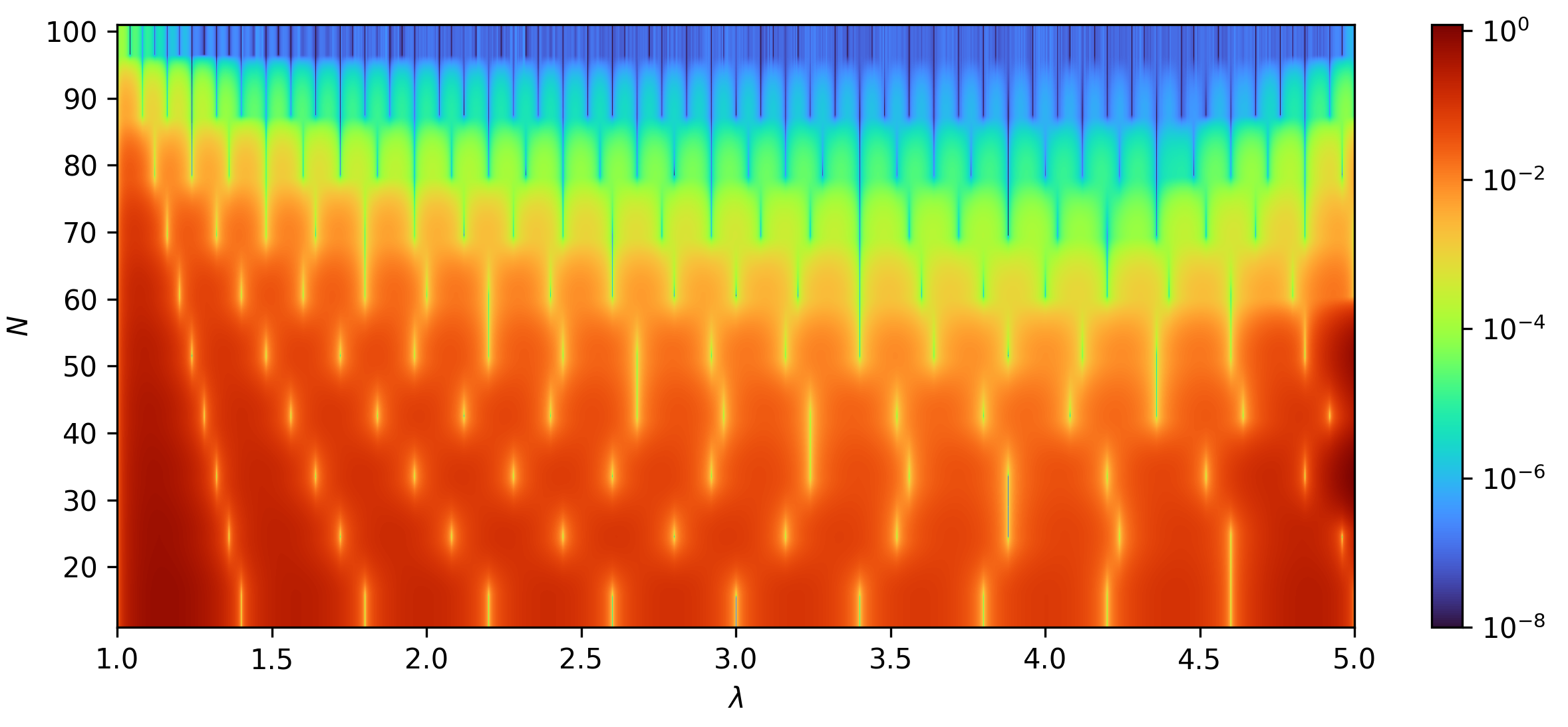}
\caption{Global errors for surrogate models built from different training discretizations $N = 11, 12, 13, 15, 17, 21, 26, 34, 51, 101$. All models are evaluated at test parameters.}
\label{fig:cmap}
\end{center}
\end{figure}

In Fig.~\ref{fig:cmap} we built a colormap of errors for different discretizations and parameter values. Naturally, the biggest errors correspond to very sparse training sets and fall below $10^{-4}$ for discretizations $\gtrsim 50$. For a specific model (an horizontal line in the colormap), bright-dark patterns describe the behavior of the model when it alternates between in-sample and out-of-sample parameter evaluations. The lowest errors usually correspond to in-sample evaluations, where splines become exact. The largest ones usually correspond to out-of-sample evaluations, where the errors due to parametric fits become relevant.

\begin{figure}[H]
\begin{center}
\includegraphics[width=1.\linewidth]{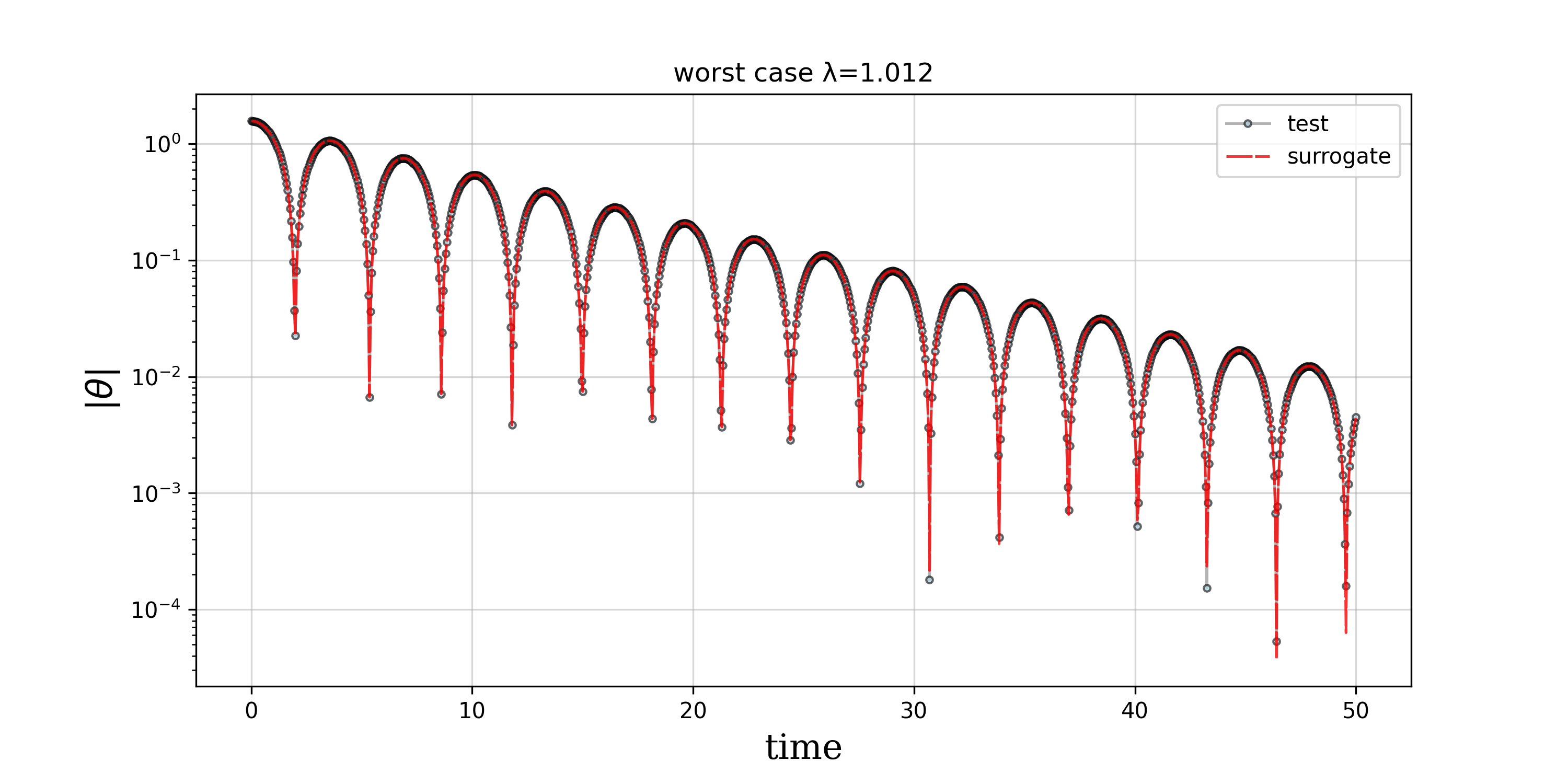}
\includegraphics[width=1.\linewidth]{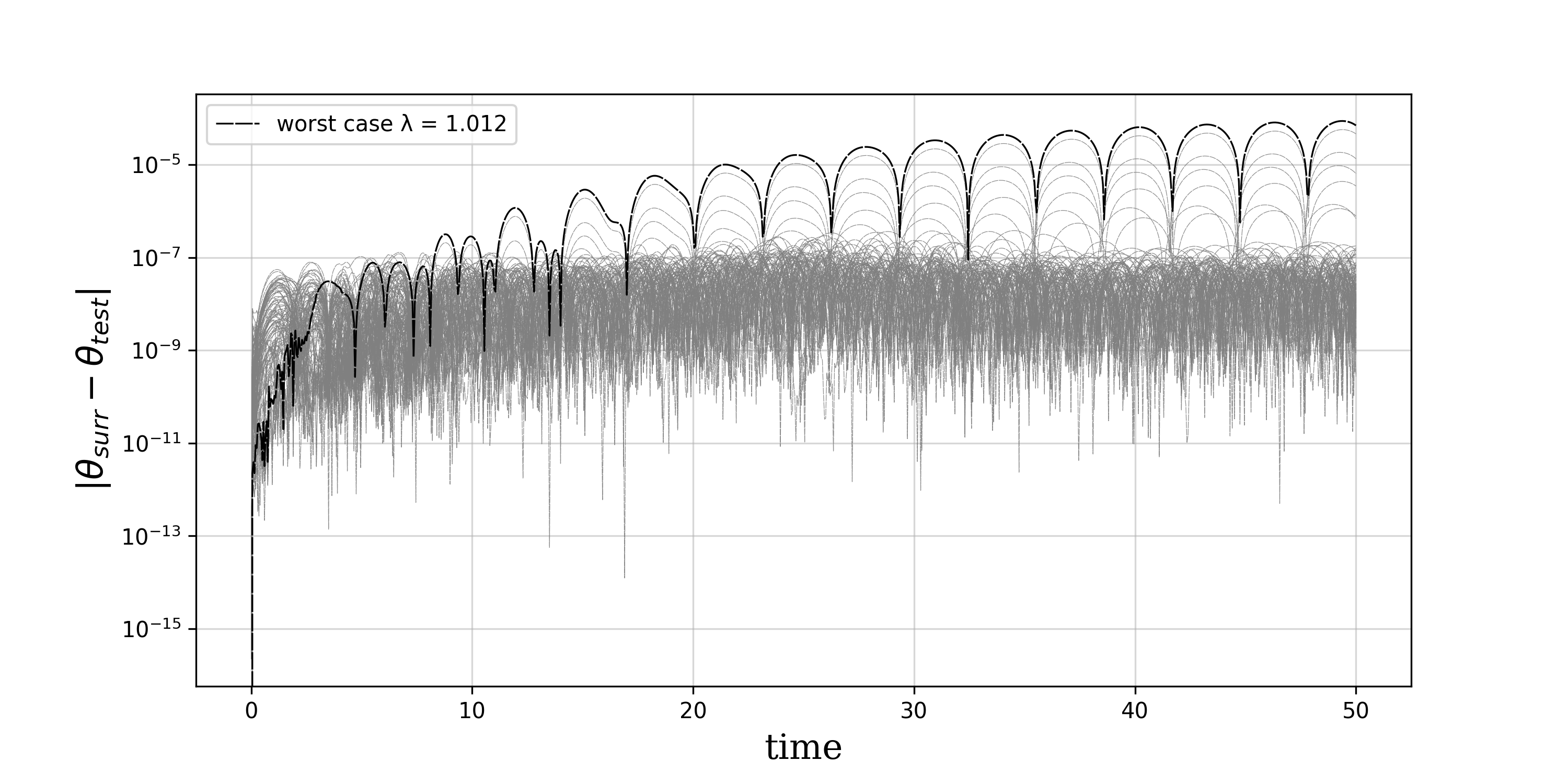}
\caption{{\bf Top}. Absolute value of function curves for both models, surrogate and ground truth. They correspond to the worst prediction in the test set. {\bf Bottom}. Point-wise difference errors for test parameters. Training points are excluded for which errors are much smaller than pure test errors.}
\label{fig:pend_errs}
\end{center}
\end{figure}

For the surrogate trained with $101$ solutions, in Fig.~\ref{fig:pend_errs} (top panel) we show the function curves for both, surrogate and ground truth models, for a parameter value corresponding to the worst global error. The surrogate evaluation at this parameter is a prediction, i.e. the associated parameter do not correspond to a training one. Since both surrogate and ground truth models are indistinguishable at eyeball resolution, we plot the absolute value for both functions in logarithmic scale to locate the dissimilarity sectors between curves better. We conclude that even the worst-case scenario shows almost no difference between surrogate and ground truth solution. The bottom panel of Fig.~\ref{fig:pend_errs} shows the absolute value of the point-wise difference between surrogate and test solutions. We removed from the test set those points which correspond to training parameters in order to focus only on generalization errors of the surrogate. We see that point-wise errors jump up at most to $\sim 10^{-4}$ whereas the bulk remains close to $\sim 10^{-7}$.

\section{Cosmic Microwave Background Anisotropies: a multidimensional case}
\label{sec:cmb}

A valuable application of surrogate models could stem from its application to modeling CMB temperature and polarization anisotropies power spectra as measured by satellites like Planck  \citep{2020A&A...641A...6P}. Such power spectra have a strong dependence on the underlying cosmological model, defined by a set of cosmological parameters $[\Omega_b h^2, \Omega_m h^2, H_0, n, \tau, A_s, 10^9 A_s exp(-2\tau)]$ , here taken to be $7$-dimensional. 

Using the CAMB (Code for Anisotropies in the Microwave Background)) for each cosmological model \citep{Lewis:1999bs},  we generated their corresponding observed temperature anisotropies power spectra. We randomly sampled the cosmological parameter space using $80,000$ points; as we will see, this is sufficiently dense. The independent or physical variable here results to be the angular multipole index $\ell$, which we sampled using $3,000$ discrete points $\ell =1, \ldots, 3,000$.

\begin{figure}[H]
\begin{center}
\includegraphics[width=1.\linewidth]{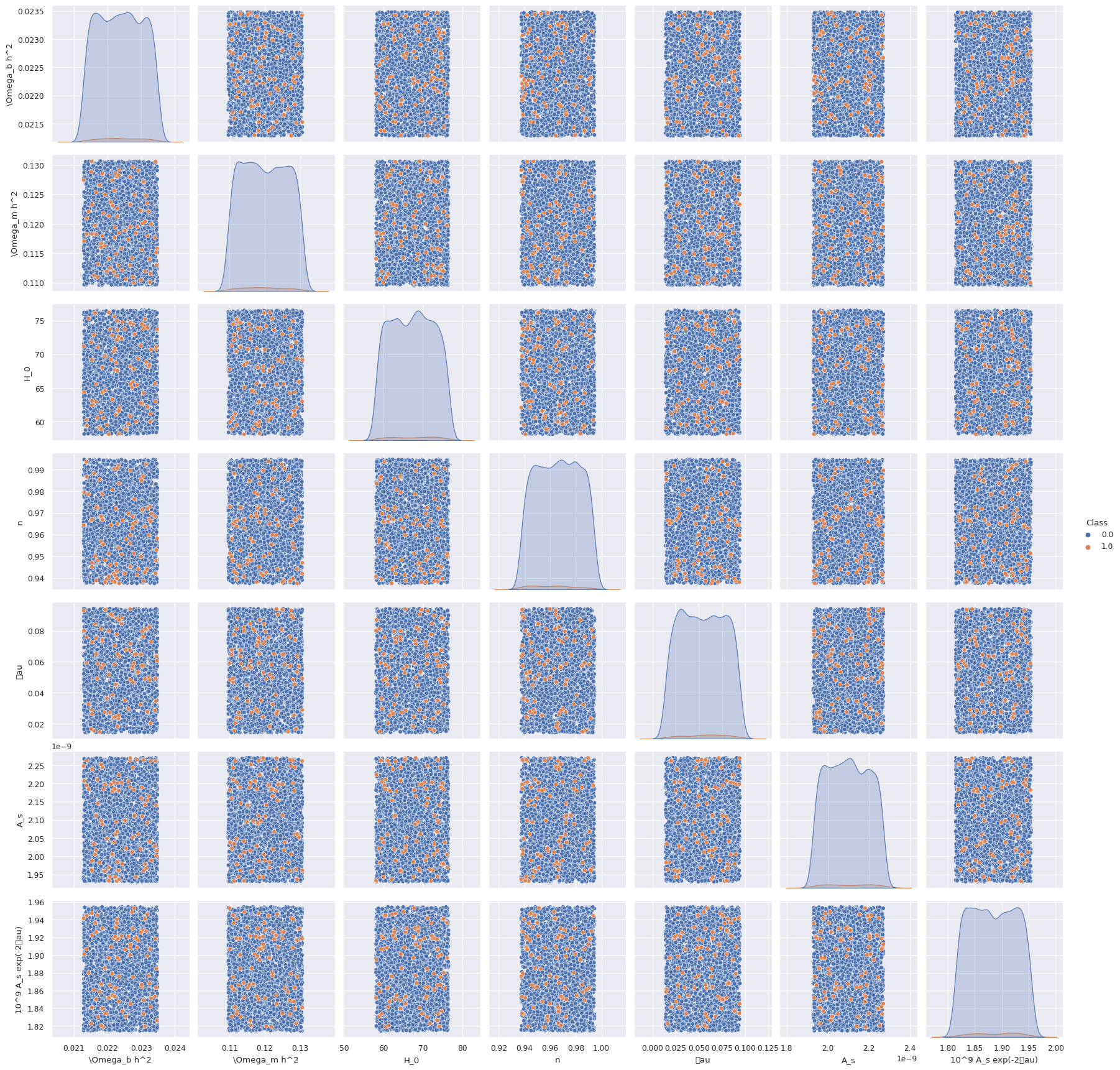}
\caption{Cosmological parameters space of a training set in blue dots and their corresponding Arby selected reduced basis set in orange points.}
\label{fig:scatter}
\end{center}
\end{figure}

We compute a reduced basis using Arby. For a greedy tolerance of $10^{-4}$ we obtain a set of greedy parameters identifying those elements in the training set that conform the reduced basis, allowing us to describe any power spectra in our sample as linear combinations of them. In particular, any training set composed of power spectra is equivalent to a set of just 84 reduced functions. Thus, for example, in Fig.~\ref{fig:scatter} we show the distribution of the cosmological parameters for a training set of $3,000$ CMB power spectra as blue points and the corresponding selected set of reduced basis as orange points. This methodology could accelerate the estimation of the cosmological parameters from CMB anisotropies by using a surrogate model built using Machine Learning algorithms on the reduced basis set. The utilization of this approach to estimate cosmological parameters will be the subject of a forthcoming publication.

\begin{figure}[H]
\begin{center}
\includegraphics[width=1.\linewidth]{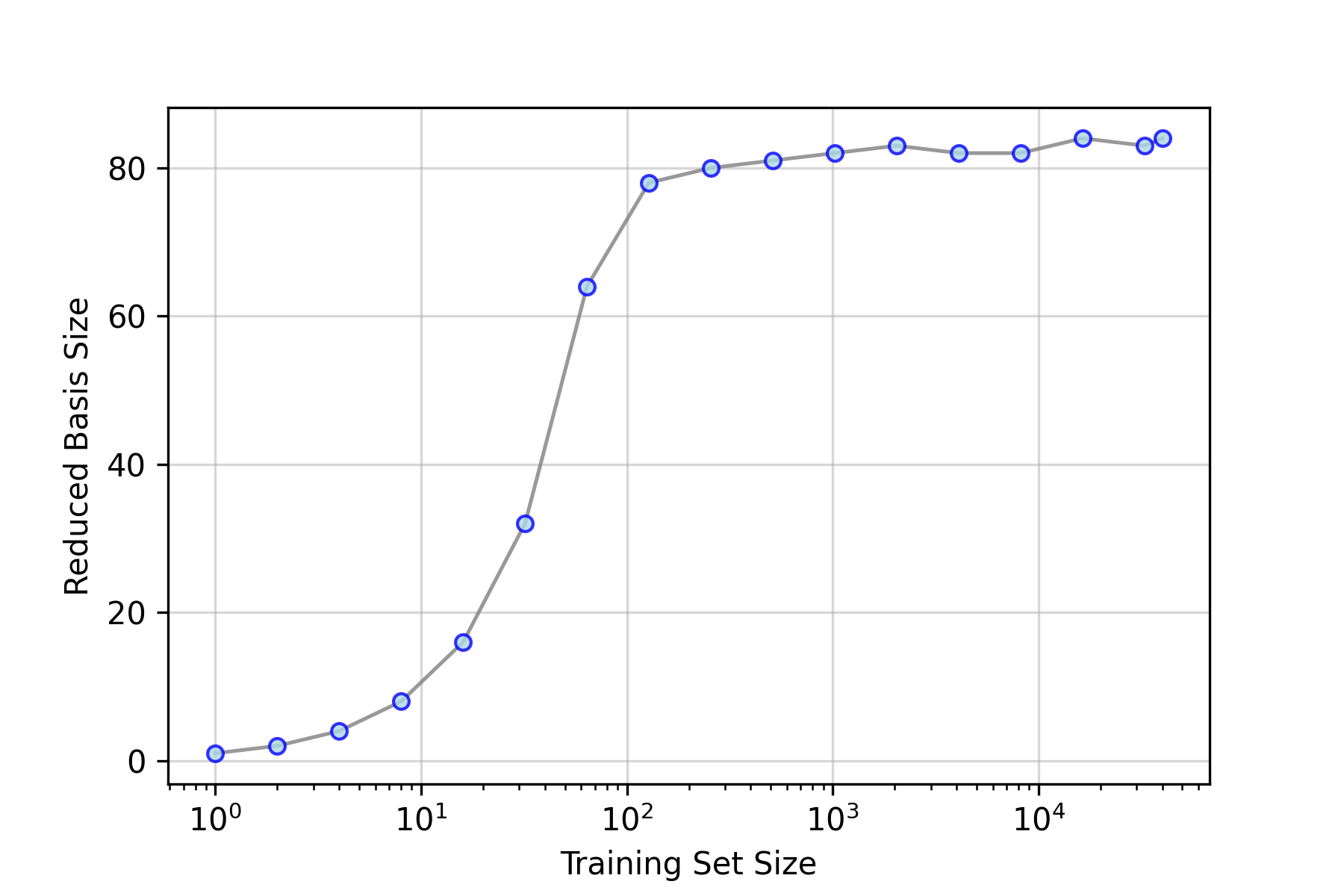}
\caption{Convergence of the number of basis elements with the size of the training set. We observe the curve saturates around $84$ basis elements.}
\label{fig:convergence}
\end{center}
\end{figure}

We report the convergence of the reduced basis number $n$ as a function of the training set size $N$ in Fig.~\ref{fig:convergence}. This plot shows that the reduced basis number $n$ stabilizes close to $n = 84$, meaning that, for the specified greedy tolerance, the training set begins to saturate at $N\sim 10^2-10^3$. This proves that the full $80,000$-points dataset involves highly redundant information, for which only $84$ functions are enough to represent the entire set. With the reduced basis we reach a compression factor of $34$ just simply by computing projections of the training set.

\section{Conclusion}
\label{sec:conclusion}

We have introduced Arby, an open source Python package that provides a set of tools to generate and handle fast and highly-accurate surrogate models in a non-intrusive way. Arby can be used to construct continuous models from sparse data composed by functions generated perhaps by differential equations or to explore redundancies in data by dimensional reduction. The offline-online architecture of Arby allows for fast deployments of predictive models that can approximate functions which otherwise can be expensive to compute.

We assessed the package with unit testing tools and ensured it satisfies proper software quality assurance, which improves in robustness and readability of code. We also perform benchmarks measuring computation times of reduced bases along different parameter combinations.

To date, for surrogate modeling Arby works on 1-D domains for both spaces, the parametric and the physical, though it supports multidimensional parameter domains for reduced basis and empirical interpolation. In future releases we want to extend Arby to several dimensions for surrogate modeling and combine it with state-of-the-art regression methods at the fitting stages. The curse of the dimensionality present in high dimensional problems like the CMB power spectra discussed in section \ref{sec:cmb} surely become a bottleneck, so parallelized scenarios are worth to be explored in the future.

The user-friendly API of Arby expands the usability of the code to virtually anyone looking for a continuous model out from discrete data, even with little or no knowledge of ROM methods.


\begin{acknowledgements}
    The authors would like to thank to their families and friends, and also IATE astronomers and Manuel Tiglio for useful comments and suggestions.
    This work was partially supported by the Consejo Nacional
    de Investigaciones Cient\'ificas y T\'ecnicas (CONICET, Argentina).
    A.V., J.B.C and M.Ch. are supported by a fellowship from CONICET.
    This research has made use of the
    \url{http://adsabs.harvard.edu/}, Cornell University xxx.arxiv.org repository, adstex (\url{https://github.com/yymao/adstex}) and the Python programming language.
\end{acknowledgements}

\bibliographystyle{aa}
\bibliography{main.bib}


\blank
\par\noindent\rule{8cm}{0.4pt}
\blank

\appendix

\section{Build reduced bases}\label{sec:rbalg}
By defining 
\begin{itemize}
\item the training set $\cK := \{ h_{\lam_i} \}_{i=1}^N$,

\item the projection error at parameter $\lam$
$$
\sigma_i(\lambda):= \| h_{\lam} - {\cal P}_{i} h_{\lam} \|^2\,,
$$
where ${\cal P}_i$ is the projector operator associated to a $i$-sized basis,

\item $\epsilon$: the greedy tolerance,

\item {\tt GS}$(h, {\tt basis})$: Orthonormalizes $h$ against a {\tt basis} through a Gram-Schmidt procedure,

\item {\tt gp}: the greedy points, and

\item {\tt rb}: the reduced basis,
\end{itemize}
the Reduced Basis-greedy algorithm proceeds as follows: 
\begin{algorithm}[H]
\caption{RB greedy algorithm}
\label{alg:Greedy}
\begin{algorithmic}[1]

\State {\bf Input:} $\cK$,  $\epsilon$ 

\vskip 10pt

\State {\bf Seed choice} (arbitrary):  $\Lam_1 \in {\cal T}$
\State $e_1 = h_{\Lam_1}/\|h_{\Lam_1}\|$
\State {\tt rb} = $\{ e_1 \}$, {\tt gp} = $\{ \Lam_1 \}$
\State $\Lam_2 = \text{argmax}_{\lam\in\cT} \sigma_1 (\lambda)$
\State $\sigma_1 = \sigma_1 (\Lambda_{2})$
\State Initialize $i=1$
\While{$\sigma_{i} > \epsilon$}
\State $i=i+1$
\State {\tt gp} = {\tt gp} $\cup \{ \Lam_i \}$
\State $e_i = {\tt GS}(h_{\Lam_i}, {\tt rb})$
\State {\tt rb} = {\tt rb} $\cup \{ e_i \}$
\State $\Lam_{i+1} = \text{argmax}_{\lam\in\cT} \sigma_i (\lambda)$
\State $\sigma_i = \sigma_i (\Lam_{i+1})$
\EndWhile
\vskip 10pt
\State {\bf Output:} {\tt rb} = $\{ e_i \}_{i=1}^n$ and {\tt gp} = $\{ \Lam_i \}_{i=1}^n$
\end{algorithmic}
\end{algorithm}

\section{Build Empirical Interpolants}\label{sec:eimalg}

\begin{algorithm}[H]
\caption{EIM algorithm}
\label{alg:EIM}
\begin{algorithmic}[1]
\State {\bf Input:} ${\tt rb} = \{ e_i \}_{i=1}^n$
\vskip 10pt
\State $X_1 = \text{argmax}_x | e_1|$ 
\For{$j = 2 \to n$} 
\State Build ${\cal I}_{j-1} [e_j](x)$
\State $r(x) = e_j(x)-{\cal I}_{j-1} [e_j](x)$
\State $X_j = \text{argmax}_x |r|$
 \EndFor
\vskip 10pt
\State {\bf Output:} EIM nodes $\{ X_i \}_{i=1}^n$ and interpolant $\cI_n$
\end{algorithmic}
\end{algorithm}

\section{On computing projection coefficients}\label{sec:stability}

The most relevant step in terms of computational cost in the RB greedy algorithm is the computation of projection coefficients, that is, step $13$ of Alg.~\ref{alg:Greedy}.
Taking full advantage of the reduced basis orthonormality $$
\langle e_i, e_j \rangle = \delta_{ij}\quad i,j=1,\ldots,n
$$
we can write projection errors as
\eq{
\sigma_n(\lam) = \| h_\lam - {\cal P}_{n} h_\lam \|^2
         = \|h_\lam\|^2 - \sum_{i=1}^n |c_i(\lam)|^2\,,
}
where $c_i(\lam) = \langle e_i, h_\lam \rangle$ are the projection coefficients. Note that
$$
\sigma_{n+1} = \sigma_{n} - |c_{n+1}|^2\,.
$$
We omitted the $\lam$ label for simplicity. This allows for constant computational cost in the addition of a new element to the basis, since one only need to compute the projection coefficients for the $n+1$ basis element while storing those corresponding to the previous basis.

In practice, the orthonormalization of the basis is not perfect and carries some error due to machine precision $\epsilon$. Therefore, inner products between basis elements write as
$$
\langle e_i, e_j \rangle = \delta_{ij} + \epsilon\,,
$$
and this error propagates through projection error as
$$
\sigma_n = \|h\|^2 - \sum_{i=1}^n |c_i|^2 +
           \epsilon \sum_{i,j=1}^n \bar{c}_i c_j \,.
$$

Then, naive implementations of this rule (saving projection coefficients to compute the next projection error) can lead to undesired error amplifications whenever $|c_i|>1$ and therefore wrong estimations of projection errors.

This is avoided simply by normalizing the training set. By doing so, we ensure $|c_i|\leq 1$ and keep controlled all orthonormalization errors.

Arby allows for normalizing options in the \code{reduced_basis} function. Depending on whether the training set is normalized or not, the computation of the greedy projection errors is done differently in each case, guaranteeing the conditioning of the algorithm at all resolutions within machine precision.

\end{document}